\documentstyle[12pt]{article}
\def\l#1#2{\raisebox{.2ex}{$\displaystyle
  \mathop{#1}^{{\scriptstyle #2}\rightarrow}$}}
\def\r#1#2{\raisebox{.2ex}{$\displaystyle
 \mathop{#1}^{\leftarrow {\scriptstyle #2}}$}}

\makeatletter
\def\eqnarray{\stepcounter{equation}\let\@currentlabel=\theequation
\global\@eqnswtrue
\global\@eqcnt\z@\tabskip\@centering\let\\=\@eqncr
$$\halign to \displaywidth\bgroup\@eqnsel\hskip\@centering
  $\displaystyle\tabskip\z@{##}$&\global\@eqcnt\@ne
  \hfil$\displaystyle{\hbox{}##\hbox{}}$\hfil
  &\global\@eqcnt\tw@ $\displaystyle\tabskip\z@
  {##}$\hfil\tabskip\@centering&\llap{##}\tabskip\z@\cr}
\@addtoreset{equation}{section}
  \def\theequation{\thesection.\arabic{equation}}
\makeatother

\begin{document}

\renewcommand{\thefootnote}{\fnsymbol{footnote}}
\newpage
\setcounter{page}{0}
\pagestyle{empty}
\begin{flushright}
{August 1996}\\
{JINR E2-96-306}\\
{hep-th/9608166}
\end{flushright}
\vfill

\begin{center}
{\LARGE {\bf Two-dimensional superintegrable }}\\[0.3cm]
{\LARGE {\bf mappings and integrable }}\\[0.3cm]
{\LARGE {\bf hierarchies in the $(2|2)$ superspace }}\\[1cm]

{\large A.N. Leznov$^{a,1}$ and A.S. Sorin$^{b,2}$}
{}~\\
\quad \\
{\em {~$~^{(a)}$ Institute for High Energy Physics,}}\\
{\em 142284 Protvino, Moscow Region, Russia}\\
{\em {~$~^{(b)}$ Bogoliubov Laboratory of Theoretical Physics, JINR,}}\\
{\em 141980 Dubna, Moscow Region, Russia}~\quad\\

\end{center}

\vfill

\centerline{ {\bf Abstract}}

The formalism of integrable mappings is applied to the problem of
constructing hierarchies of $(1+2)$ dimensional integrable systems in
the $(2|2)$ superspace. We find new supersymmetric integrable
mappings and corresponding to them new hierarchies of integrable
systems which, at the reduction to the $(1|2)$ superspace, possess
$N=2$ supersymmetry. The general formulae obtained for the
hierarchies are used to explicitly derive their first nontrivial
equations possessing a manifest $N=2$ supersymmetry. New bosonic
substitutions and hierarchies are obtained from the supersymmetric
counterparts in the bosonic limit.

\vfill
{\em E-Mail:\\
1) leznov@mx.ihep.su\\
2) sorin@thsun1.jinr.dubna.su }
\newpage
\pagestyle{plain}
\renewcommand{\thefootnote}{\arabic{footnote}}
\setcounter{footnote}{0}

\section{Introduction}

The method of discrete substitutions is one of the shortest
and most direct ways of describing and solving equations of integrable
systems \cite{1}, \cite{2}.

The integrable mappings and their properties play the key role
in this approach. All other ingredients of the theory, such as
the explicit form of integrable hierarchies and the solution of the
corresponding systems of equations, are a direct corollary
of the representation theory of the discrete group of integrable mappings.

The discrete substitution is the same for each system of equations
in the given integrable hierarchy and thus contains detailed information
about all of the sets of integrable systems belonging to it.

Of course, it is necessary to keep in mind that up to now there is no
rigorous, from a mathematical point of view, classification theory of
discrete substitutions themselves nor a representation theory of
the group of integrable mappings. Nevertheless, if under
some other consideration it is possible to obtain an explicit form
of a discrete substitution, all other corollaries may be
obtained by straightforward calculations.

The algorithm of these calculations is very simple and resembles a
computer program: it is necessary to perform many identical
operations that can be interrupted at an arbitrary step
and thus obtain relevant information about some system of equations
belonging to the hierarchy.

The goal of the present letter is to apply this approach to the problem
of constructing two-dimensional supersymmetric hierarchies of integrable
systems. We demonstrate this approach by
supersymmetric integrable mappings connected with the two-dimensional
supersymmetric Toda lattice.
It is also possible to consider this construction as one of the
possible generalizations of the Darboux transformation to the
supersymmetric case.

\section{Supersymmetric substitutions in the $(2|2)$ superspace}

Below, we briefly discuss the main points of the discrete
transformational approach, taking as the simplest examples, the
integrable supersymmetric substitutions related to the
supersymmetric Toda lattice in two dimensions.

It is well known that under appropriate boundary conditions, the
supersymmetric Toda lattice
\begin{equation}
D_-D_+ \ln b=\r {b}{}-\l {b}{} \label{1}
\end{equation}
is an exactly integrable system (see e.g., \cite{3} and references
therein). Here $b$ is the bosonic superfield
$b=b^0+\theta_+b^-+\theta_-b^+ + \theta_+\theta_- b^2$; $b^0,b^2$
($b^-,b^+$) are bosonic (fermionic) fields, and $\theta_+,\theta_-$ are
elements of the  Grassman algebra. The $D_{\pm}$ are the $N=1$
supersymmetric fermionic covariant derivatives \begin{eqnarray}
 D_+=\frac{\partial}{\partial \theta_+}+\theta_+{\partial}_x,
\quad D_-=\frac{\partial}{\partial \theta_-}+\theta_-{\partial}_y, \quad
 D^2_+={\partial}_x,\quad
 D^2_-={\partial}_y,\quad \{D_+,D_-\}=0,
\end{eqnarray}
where $x$ and $y$ are two independent space coordinates; the notation
$\r {b}{} (\l {b}{})$ means that the index of variable $b$ is shifted
by $+1$ ($-1$) (for instance, $\r {b}{}_n\equiv b_{n+1}$, and so on).

Now let us introduce the fermionic $f$, or bosonic ${\bar b}$, superfields
by the following chain of relations:
\begin{equation}
\r {f}{}=f+D_+ \ln \r {b}{} \label{ee}
\end{equation}
or
\begin{equation}
{\bar b}=\r {b}{} \label{be}
\end{equation}
and using (\ref{ee}) or (\ref{be}), identically rewrite the set of Toda
lattice equations (\ref{1}) as
\begin{equation}
\r {b}{}=D_-f-b,\quad \r {f}{}=f+D_+ \ln \r {b}{} \label{2}
\end{equation}
or
\begin{equation}
\r {{\bar b}}{}=b+D_-D_+\ln {\bar b},\quad \r {b}{}={\bar b}, \label{2b}
\end{equation}
respectively.
The relation (\ref{2}) ((\ref{2b})) can be considered as a definition of
some mapping: it is a rule used to associate two arbitrary initial
functions $f,b$ (${\bar b},b$) with the final ones $\r {f}{},\r {b}{}$
($\r {\bar b}{},\r {b}{}$). It is not difficult to check that the
mappings\footnote{We also call them substitutions.}
(\ref{2}) and (\ref{2b}) are invertible and the inverse transformations
have the form
\begin{equation}
\l {b}{}= D_-\l {f}{} - b,\quad \l {f}{}=f-D_+ \ln b \label{3}
\end{equation}
or
\begin{equation}
{\bar b}=\l {b}{}+D_-D_+ \ln \l {\bar b}{},\quad b=\l {\bar b}{}. \label{3b}
\end{equation}

Substitutions (\ref{2}) and (\ref{2b}) possess the inner automorphisms
$\sigma$ with the properties
\begin{eqnarray}
\sigma=\sigma_2 \sigma_1, \quad
&&\sigma_1 b {\sigma}^{-1}_1=-b, \quad
\sigma_1 f {\sigma}^{-1}_1=D_+D^{-1}_-f, \quad
\sigma_1 D_+{\sigma}^{-1}_1=D_+ , \quad
\sigma_1 D_-{\sigma}^{-1}_1=D_-,
\nonumber\\
&&\sigma_2 b {\sigma}^{-1}_2=b,    \quad
\sigma_2 f {\sigma}^{-1}_2=f,    \quad
\sigma_2 D_+{\sigma}^{-1}_2=D_-, \quad
\sigma_2 D_-{\sigma}^{-1}_2=D_+;
\label{auto1}
\end{eqnarray}
\begin{eqnarray}
\sigma b {\sigma}^{-1}=-b, \quad
\sigma {\bar b} {\sigma}^{-1}=-{\bar b}, \quad
\sigma D_+{\sigma}^{-1}=D_- , \quad
\sigma D_-{\sigma}^{-1}=D_+,
\label{auto2}
\end{eqnarray}
respectively, which will be usefull in what follows.
The action of $\sigma$ on the covariant derivatives $D_{\pm}$ can be
induced by the following transformation of the $(2|2)$ superspace
coordinates
\begin{eqnarray}
\sigma x {\sigma}^{-1}=y, \quad
\sigma y {\sigma}^{-1}=x, \quad \sigma {\theta}_
+ {\sigma}^{-1}={\theta}_- , \quad
\sigma {\theta}_- {\sigma}^{-1}={\theta}_+ .
\label{auto}
\end{eqnarray}

The next substitution connected with the super Toda lattice (\ref{1})
is the literal generalization of the Darboux transformation (direct and
inverse) to the supersymmetric case
\begin{eqnarray}
\r{u}{}&=&\frac{1}{v},\quad
\r{v}{}=v(D_-D_+\ln v-uv);
\nonumber\\
\l{v}{}&=&\frac{1}{u},\quad
\l{u}{} =-u(D_-D_+\ln u +uv),
\label{6}
\end{eqnarray}
where $u$ and $v$ are bosonic superfields. Substitution (\ref{6})
possess the global $U(1)$-invariance. With respect to the U(1) group, the
superfields $u$ and $v$ have opposite $U(1)$-charges. From (\ref{6}), it
immediately follows that the function $T_0=uv$ satisfies the equation for
the supersymmetric Toda lattice (\ref{1}):
\begin{equation} D_-D_+\ln T_0 =\r {T_0}{}-\l {T_0}{}.
\label{7}
\end{equation}

The most important object for investigations closely connected with the
substitution, and directly following from it, is the symmetry equation.
Following \cite{1}, \cite{2}, let us recall that the symmetry
equation for a given substitution can be obtained by differentiation of
the substitution with respect to an arbitrary independent argument, or
parameter. Denoting a derivative from all functions involved in the
substitution by new letters and considering them as independent
functionals whose arguments are the above-mentioned functions and their
superspace derivatives, one can find a set of equations for these
functionals which are called the symmetry equations. If a symmetry
equation possesses nontrivial solutions and it is possible to construct
them, then one can produce an evolution-type system of integrable
equations by a simple algorithmic procedure using only these solutions.
Substitutions of this kind have been called integrable (for details,
see \cite{1}, \cite{2}).

In the case under consideration, the independent arguments of a
substitution are the coordinates of the $(2|2)$ superspace. Thus, as
different from bosonic coordinate space, in the case of superspace, it is
possible to differentiate with respect to its even or odd coordinates
(parameters). As a consequence, knowledge of the symmetry equation
solutions allows us to produce the evolution-integrable equations
with either even or odd evolution parameters which, nevertheless,
belong to the same integrable hierarchy. Here we discuss only the
former case.

To illustrate the above-general discussion, for definiteness, we restrict
ourselves to a concrete example of the substitution (\ref{2}); however,
at the end of the next section, we present the results of calculations
for all of the substitutions discussed in this section.

Using the general rules described above, we obtain the following symmetry
equation corresponding to the substitution (\ref{2}):
\begin{equation}
\r{B}{}=D_-F-B,\quad
\r{F}{}=F+D_+ ({\r{B}{}\over {\r{b}{}}}{}).
\label{4}
\end{equation}
Here, $F$ and $B$ are fermionic and bosonic functionals whose independent
arguments are $ f,b,D_{\pm}f,D_{\pm}b,b_x,b_y, f_x,f_y,\ldots$ . The
functionals $\r {B}{s}, \l {B}{s},\r {F}{s},\l {F}{s}$ are the same
functionals whose arguments are shifted s-times in the direct or
inverse direction and are connected with the initial arguments $b,f$ by
relations (\ref{2}) and (\ref{3}).

By construction, the pair $F=f',B=b'$ (the sign ${}'$ means differentiation
with respect to each of the independent bosonic arguments of the problem)
is a solution of eq. (\ref{4}). This solution was called a trivial one in
\cite{1}, \cite{2} and, in this sense, this term has been used
above.

Each solution of the symmetry equation (\ref{4}) is connected with the
evolution-type integrable system
\begin{equation}
b_t=B,\quad f_t=F. \label{5}
\end{equation}
Moreover, the last one is invariant with respect to the discrete
transformation of the substitution (\ref{2}), and the symmetry equation
(\ref{4}) is exactly the condition of this invariance.

The hierarchies of integrable systems are encoded in integrable
substitutions and can be explicitly obtained by solving their symmetry
equations. In the next section, we give an infinite set of partial
solutions of the symmetry equation for substitutions
(\ref{2}), (\ref{2b}), and (\ref{6}).

We would like to close this section with a few remarks.

First, all substitutions in the $(2|2)$ superspace can be reduced to the
$(1|2)$ superspace. In this case, all of the considered functions must
have a dependence on only three arguments: $x+y$, $\theta_+$, and
$\theta_-$.  At this reduction, the substitutions (\ref{2b}) and (\ref{6})
become $N=2$ supersymmetric substitutions. This statement becomes evident
if one takes into account that the fermionic covariant derivatives
$D_{\pm}$ enter (\ref{2b}) and (\ref{6}) only in the $N=2$
superinvariant combination $D_-D_+$ (for details, see section 4).  It is
easily to understand that the same statement is correct with respect
to the substitution (\ref{2}) after introducing the new bosonic superfield
${\tilde b}$ by the relation \begin{equation} f=D_+{\tilde b}. \label{bb}
\end{equation}
Indeed, in terms of the superfields $b,{\tilde b}$, the substitution
(\ref{2}) has the following desirable form:
\begin{equation}
\r{b}{}=D_-D_+{\tilde b}-b,\quad
\r {{\tilde b}}{}= {\tilde b}+\ln \r {b}{}
\label{b}
\end{equation}
up to the possible unessential constant on the right hand
side of the second relation. It is interesting to note that up to this
constant, the substitution (\ref{b}) possesses the local inner
automorphism $\sigma$
\begin{eqnarray}
\sigma b {\sigma}^{-1}=-b, \quad
\sigma {\tilde b} {\sigma}^{-1}={\tilde b},\quad
\sigma D_+{\sigma}^{-1}=D_-, \quad
\sigma D_-{\sigma}^{-1}=D_+,
\label{auto3}
\end{eqnarray}
as different from substitution (\ref{2}), possessing the
nonlocal automorphism (\ref{auto1}).

Second, the $N=2$ superinvariance of the substitutions (\ref{2b}),
(\ref{6}) and (\ref{b}) guarantees the same invariance for the all
integrable hierarchies related to these substitutions.

\section{Solution of the symmetry equation}

In the case of usual space, the problem of the title of this
section was solved in \cite{4}. An algorithm
for recurrent calculations has been proposed which allows one to pass
step by step. This process may be interrupted at an arbitrary
step for which at least one obvious simple solution always exists.
Here, we generalize these calculations to the case of superspace.
One essential difference, compared to the case of usual
space, consists of the fact that in superspace, the recurrent
procedure can be interrupted only at an even step but not at
an odd one.

For definiteness, we restrict ourselves to the substitution (\ref{2}).

 From the symmetry equation (\ref{4}), it immediately follows that
an unknown bosonic function may be represented in the form
\begin{eqnarray}
{B\over b}=\alpha_0-\l {\alpha_0}{} ,
\end{eqnarray}
and a new unknown bosonic function $\alpha_0$ is the solution of the equation
\begin{equation}
D_-D_+\alpha_0=\r {b}{}(\r {\alpha_0}{}-\alpha_0)+
b(\alpha_0-\l {\alpha_0}{}). \label{8}
\end{equation}

In terms of the solution to eq.(\ref{8}), the evolution-type
integrable system (\ref{5}) can be rewritten as
\begin{equation}
b_t=b({\alpha_0}-\l {\alpha_0}{}),\quad
f_t=D^{-1}_-[\r {b}{}(\r {\alpha_0}{}-\alpha_0)+
b(\alpha_0-\l {\alpha_0}{})].  \label{9}
\end{equation}

Now let us describe the recurrent steps for the solution of the symmetry
equation (\ref{8}), linear with respect to an unknown function
${\alpha_0}$.

The symmetry equation, together with the equation for the bosonic
function $b$ (which arises after excluding the fermionic function $f$
 from the system (\ref{2})), may be rewritten in an equivalent form
suitable for further calculations:
\begin{equation}
D_+ \alpha_0=D^{-1}_-[\r {b}{}(\r {\alpha_0}{}-\alpha_0)+
b(\alpha_0-\l {\alpha_0}{})],\quad
D_+ b=b D^{-1}_-(\r {b}{}-\l {b}{}). \label{10}
\end{equation}
Simple inspection of the eqs.(\ref{10}) shows that they possess the inner
automorphism $\sigma$ with the properties
\begin{eqnarray}
\sigma {\alpha_0} {\sigma}^{-1}={\eta_0}, \quad
\sigma b {\sigma}^{-1}=-b, \quad
\sigma D_+{\sigma}^{-1}=D_- , \quad
\sigma D_-{\sigma}^{-1}=D_+,
\label{auto4}
\end{eqnarray}
where ${\eta_0}$ is another solution of the eq.(\ref{8}). We
use this automorphism to construct one-parametric family of
solutions of the eq.(\ref{8}) (see the end of this section).

We present below a series of straightforward,
simple transformations with short comments. First, let us introduce the
new fermionic function ${\tilde \alpha_0}$:
$$
\alpha_0=D^{-1}_-{\tilde \alpha_0}.
$$
Keeping in mind that the operator $D_-$ is odd and the
corresponding rules of working with such objects, we come to the
equation for ${\tilde \alpha_0}$:
\begin{equation}
-D_+ {\tilde \alpha_0}=\r {b}{}D^{-1}_-
(\r {{\tilde \alpha_0}}{}-{\tilde \alpha_0})+
bD^{-1}_-({\tilde \alpha_0}-\l {{\tilde \alpha_0}}{}). \label{11}
\end{equation}
Second, we use the following substitution:
$$
{\tilde \alpha_0}=\r {b}{}\alpha_1+b \beta_1,
$$
and, as a result, we get the corresponding system of equations for the
unknown functions $\alpha_1,\beta_1$,
\begin{eqnarray}
&-&D_+ \alpha_1 + {\alpha}_1 D^{-1}_-(\r {b}{2}-b)=
D^{-1}_-(\r {{\tilde \alpha_0}}{}-{\tilde \alpha_0}),
\nonumber\\
&-&D_+ \beta_1 + {\beta}_1 D^{-1}_-(\r {b}{}-\l {b}{})=
D^{-1}_-({\tilde \alpha_0}-\l{{\tilde \alpha_0}}{}) ,
\label{12}
\end{eqnarray}
after equating the coefficient functions to zero at the $\r {b}{},b$ terms.
Let us stress that this is an additional independent assumption.
Subtracting the second equation, shifted by one step to the left, from
the first, we obtain the following equation:
$$
-D_+(\alpha_1- \r {\beta}{}_1) +(\alpha_1-\r {\beta}{}_1) D^{-1}_-
(\r {b}{2}-b)=0.
$$
 From the last equation, we see that the system (\ref{12}) possesses a
partial solution for which $\alpha_1=\r {\beta_1}{}$. In what follows,
we work exactly with a solution of this kind, and (\ref{12}), in
this case, is equivalent to a single equation for the unknown
function $\alpha_1$:
\begin{equation}
-D_+ \alpha_1 +{\alpha}_1 D^{-1}_-(\r {b}{2}-b)=
D^{-1}_-(\r {b}{2}\r {\alpha_1}{}-
b \l {\alpha_1}{}). \label{13}
\end{equation}
 From this place, it is necessary to repeat the circle of calculations
mentioned in the introduction. Thus, after introducing the new
function ${\tilde {\alpha}_1}$ in the following way:
\begin{equation}
\alpha_1= D^{-1}_-{\tilde {\alpha}_1}
\end{equation}
the equation for ${\tilde {\alpha}_1}$ takes the form
\begin{equation}
D_+ {\tilde {\alpha}_1} +{\tilde {\alpha}_1} D^{-1}_-(\r {b}{2}-b)=
\r {b}{2}D^{-1}_-(\r {\tilde {\alpha}_1}{}+{\tilde {\alpha}_1})-
bD^{-1}_-({\tilde {\alpha_1}}+\l {\tilde {\alpha_1}}{}). \label{14}
\end{equation}
Let us notice that the function ${\tilde {\alpha}_1}$ is bosonic in
contrast to its first-step counterpart ${\tilde {\alpha}_0}$.
After the substitution
\begin{equation}
{\tilde {\alpha}_1}=\r {b}{2}\alpha_2+b \beta_2 ,
\end{equation}
we come to a system of two equations, in the same way as to (\ref{12}),
for two unknown functions $\alpha_2,\beta_2$, which now
possesses the partial solution $\alpha_2=-\r {\beta_2}{}$ to be
utilized. The single equation for the unknown function $\alpha_2$
takes the form
\begin{equation}
D_+ \alpha_2+{\alpha}_2 D^{-1}_-(\r {b}{3}-\r {b}{}+\r {b}{2}-b)=
D^{-1}_-(\r {b}{3}\r {\alpha_2}{}-\r {b}{}\alpha_2+\r {b}{2} \alpha_2-
b \l {\alpha_2}{}). \label{15}
\end{equation}
In contrast to the analogous equation (\ref{13}), eq. (\ref{15})
possesses the partial solution $\alpha_2=1$. By this solution, it is
possible to interrupt the calculation procedure or continue it according to
the proposed scheme. As a result, we obtain the following recurrent
relations:
\begin{eqnarray}
\alpha_{2n}=D^{-1}_-(\r {b}{2n+1} \alpha_{2n+1}
+ b \l {\alpha}{}_{2n+1}{}), \quad
\alpha_{2n+1}=D^{-1}_-(\r {b}{2n+2} \alpha_{2n+2}
- b \l {\alpha}{}_{2n+2}{}), \label{16}
\end{eqnarray}
where $\alpha_{2n}$ ($\alpha_{2n+1}$) are bosonic (fermionic) functions
and, for every even step, there is a partial solution $\alpha_{2n}=1$.

Applying the automorphism transformations (\ref{auto4}) to the recurrent
relations (\ref{16}), we get the new recurrent relations
\begin{eqnarray}
\eta_{2n}=-D^{-1}_+(\r {b}{2n+1} \eta_{2n+1}
+ b \l {\eta}{}_{2n+1}{}), \quad
\eta_{2n+1}=-D^{-1}_+(\r {b}{2n+2} \eta_{2n+2}
- b \l {\eta}{}_{2n+2}{}), \label{16a}
\end{eqnarray}
generating another solution ${\eta_0}$ to the eq.(\ref{8}).

In terms of these two solutions of the symmetry equation (\ref{8}), for
the substitution (\ref{2}), the integrable system of evolution equations
(\ref{9}) can be represented in the following form:
\begin{equation}
b_t=b({\gamma_0}-\l {\gamma_0}{}),\quad
f_t=D^{-1}_-[\r {b}{}(\r {\gamma_0}{}-\gamma_0)+
b(\gamma_0-\l {\gamma_0}{})],  \label{9a}
\end{equation}
where
\begin{equation}
{\gamma_n}\equiv{\alpha_n}+h{\eta_n}
\label{9aa}
\end{equation}
and $h$ is an arbitrary parameter\footnote{The coefficient at ${\alpha_0}$
in (\ref{9a}), (\ref{9aa}) is ineffective and it is always possible to put
it equal to unity by the corresponding rescaling of the evolution variable
$t$.}.

Thus, the expressions (\ref{16})-(\ref{9a}) give us
explicit formulae for the one-parametric hierarchy of integrable equations
corresponding to the integrable substitution (\ref{2}).
In what follows, we choose the parameter $h$ equal to zero for all of the
discussed substitutions to simplify the formulae, keeping in mind that it
is always possible to restore it by the action of an inner automorphism
transformation, corresponding to a given substitution (see e.g.,
eqs.(\ref{auto1}) and (\ref{auto2}) for the substitutions (\ref{2}) and
(\ref{2b}), respectively), on the solution with $h=0$ and then adding to it
the obtained result multiplied by $h$.

To close this section, let us present the results of calculations for
the substitutions (\ref{2b}) and (\ref{6}).

For the substitution (\ref{2b}), the hierarchy of integrable equations
can be written in the following form:
\begin{equation}
b_t=b({\gamma_0}-\l {\gamma}{}_0),\quad
{\bar b}_t=\r {b}{}(\r {\gamma}{}_0-\gamma_0),  \label{17}
\end{equation}
where $\gamma_{n}$ is defined by eq.(\ref{9aa}) and $\alpha_n$ ($\eta_n$)
satisfies the same recurrent relations (\ref{16}) ((\ref{16a})).

The integrable equations of the hierarchy corresponding to the substitution
(\ref{6}) look like
\begin{equation}
u_t=uD^{-1}_-(\l {T}{}_0 \l {\gamma}{2}_1+ T_0 \l {\gamma}{}_1) ,\quad
v_t=-vD^{-1}_-(T_0 \l {\gamma}{}_1+ \r {T}{}_0 \gamma_1),  \label{18}
\end{equation}
where $\alpha_n$ and $\eta_n$ satisfy the same recurrent relations
(\ref{16}) and (\ref{16a}), respectively, in which the function $b$ is
replaced by $T_0$ (\ref{7}).

It is easy to observe that at the interruption of the recurrent relations
at the $n$-th step (i.e., at $\alpha_{2n}=\eta_{2n}=1$), the maximal
order of the linearly appearing bosonic derivative on the right-hand side
of the eqs.(\ref{9a}), (\ref{17}) and (\ref{18}) is equal to just $n$.
To use the terminology of the inverse scattering theory suitable to the
one-dimensional case, one can say that the interruption condition
$\alpha_{2n}=\eta_{2n}=1$ extracts the $n$-th flow.

\section{Examples: the $n=2$ cases and their bosonic limits}

Using the general formulae of the last section, we present here the
results of the calculations for the first nontrivial equations of the
integrable hierarchies corresponding to the substitutions
(\ref{2}), (\ref{2b}) and (\ref{6}), as well as for their bosonic limits.
These cases correspond to the interruption of the recurrent procedures
(\ref{16}) and (\ref{16a}) at the second step (at $n=2$), i.e., we put
$\alpha_{4}=\eta_{4}=1$ in eqs.(\ref{16}) and (\ref{16a}). Let us give
the results of the calculations with short comments.

{}~

{\large {\bf 4.1 Substitution (\ref{2}) }}
\begin{eqnarray}
 f_t &=& f_{xx}+2D^{-1}_-\{((D_-f)D_+f)_x-D_+[f_xD_-f-(fb)_x
 +bD^{-1}_-b_x] -b_{xx}\}
 -2D_+(fD^{-1}_-b_x),
\nonumber\\
 b_t &=& -b_{xx}+2(bD_+f)_x-2D_+(bD^{-1}_-b_x).
\label{19}
\end{eqnarray}
Under the reduction of (\ref{19}) from the superspace $(2|2)$ to
superspace $(1|2)$ (see discussion at the end of section 2), it can be
represented in the following form:
\begin{eqnarray}
 f_t &=& f_{xx}+D_+((D_+f)^{2}-2bD_-f +
 2D_-D_+b+2b^{2}), \nonumber\\
 b_t &=& -b_{xx}+2(bD_+f)_x-D_+D_-b^{2}.
\label{20}
\end{eqnarray}
As was mentioned at the end of section 2, in this case one can get
the $N=2$ supersymmetric system of equations if one introduces the
superfield ${\tilde b}$ (\ref{bb}) instead of $f$. Thus, in the new
terms, one can rewrite (\ref{20}) as
\begin{eqnarray}
 {\tilde b}_t &=& {\tilde b}_{xx}+{{\tilde b}_x}^{2}-
 ib[D,{\bar D}]{\tilde b} + i[D,{\bar D}] b + 2b^{2},
\nonumber\\
 b_t &=& -b_{xx}+2(b {\tilde b}_x)_x + \frac{i}{2} [D,{\bar D}] b^{2}
\label{21}
\end{eqnarray}
and the $N=2$ supersymmetry of (\ref{21}) becomes manifest. Here $i$ is the
imaginary unity and $D$, ${\bar D}$ are two $N=2$ supersymmetric fermionic
covariant derivatives with opposite U(1)-charges, related to their $N=1$
counterparts $D_{\pm}$ by the following formulae:
\begin{eqnarray}
 D= \frac{1}{\sqrt {2}}(D_-+iD_+), \quad  {\bar D}=
 \frac{1}{\sqrt {2}}(D_--iD_+), \quad
 \{D,{\bar D}\}=2{\partial}_x, \quad D_-D_+=\frac{i}{2}[D,{\bar D}].
\label{28}
\end{eqnarray}
Of course, the eqs.(\ref{20}) also possess the $N=2$ supersymmetry;
however, in this form, it is hidden. It is interesting that both (\ref{20})
and (\ref{21}) are local despite the fact that the $U(1)$ transformation
 from the $N=2$ supergroup is nonlocally realized for (\ref{20}).

We would next like to describe the bosonic limit of the substitution
(\ref{2}) and the eqs.(\ref{19}). We define the
bosonic components of the superfields $b,f$ as\footnote{Let us
remember that in the bosonic limit, all fermionic components
must be put equal to zero.}
\begin{equation}
b \vert=a,\quad D_-D_+b \vert=w, \quad
D_+f \vert=c,\quad D_-f \vert=d,
\label{boscom3}
\end{equation}
where $|$ means the $({\theta}_+,{\theta}_-) \to 0$ limit.
In terms of these components, the desirable expressions have the following
form:
\begin{equation}
\r {a}{}=d-a,\quad \r w{}=-c_y-w, \quad
\r{c}{}=c+({\ln {\r {a}{}}})_x,\quad \r d{}=d+\frac{{\r {w}{}}}{{\r {a}{}}}
\label{bossub3}
\end{equation}
for the substitution and
\begin{eqnarray}
 d_t &=& (d-2a)_{xx}+2cd_x+2ac_x+2(a-d){{\partial}_y}^{-1}w_x, \quad
 a_t = -a_{xx} + 2(ac)_x +2a{{\partial}_y}^{-1}w_x,
\nonumber\\
 c_t &=& [c_x + c^{2} + {{\partial}_y}^{-1}(2w+b^{2}-2ad)_x]_x,\quad
 w_t = [-(w+b^{2})_x+2wc+2ad_x]_x
\label{bos3}
\end{eqnarray}
for the eqs.(\ref{19}).

 From the substitution (\ref{2}),
as well as directly from (\ref{19})-(\ref{21}), one can see that the
scaling dimensions of all superfields are completely fixed and are defined
by the relations:  $[b]={cm}^{-1}$, $[{\tilde b}]={cm}^{0}$,
$[f]={cm}^{-1/2}$.  As a consequence, for the components (\ref{boscom3}),
we get the following values for the scaling dimensions:
$[a]=[c]=[d]={cm}^{-1}$, $[w]={cm}^{-2}$.

{}~

{\large {\bf 4.2 Substitution (\ref{2b}) }}
\begin{eqnarray}
 b_t &=& -b_{xx}-2(bD^{-1}_-D_+ {\bar b})_x-2D_+((D_+b)D^{-1}_-D_+b),
\nonumber\\
 {\bar b}_t &=& {\bar b}_{xx}-2({\bar b}D^{-1}_-D_+ b)_x-
 2D_+((D_+{\bar b})D^{-1}_-D_+{\bar b}).
\label{22}
\end{eqnarray}
At the reduction to the $(1|2)$ superspace, one can rewrite (\ref{22}) in
the following form:
\begin{eqnarray}
 b_t &=& -b_{xx}-i(b{\partial}^{-1} [D,{\bar D}] {\bar b})_x-
 ib_x{\partial}^{-1} [D,{\bar D}] b+ 2i(Db){\bar D}b,
\nonumber\\
 {\bar b}_t &=& {\bar b}_{xx} -i({\bar b}{\partial}^{-1} [D,{\bar D}] b)_x -
 i{\bar b}_x{\partial}^{-1} [D,{\bar D}]{\bar b}+
 2i(D{\bar b}){\bar D}{\bar b}.
\label{23}
\end{eqnarray}
Equations (\ref{23}) possess manifest, local $N=2$ supersymmetry;
however, they are nonlocal. Nevertheless, it is possible to localize
them.  In order to do this, let us introduce new fermionic
superfields ${\psi},{\bar {\psi}}$ in the following way:
$b=D_-{\psi}$   ${\bar b}=D_-{\bar {\psi}}$. In these new terms, the
systems (\ref{22}), (\ref{23}) have the form
\begin{eqnarray}
{\psi}_t &=&
 -{\psi}_{xx}+2D^{-1}_-((D_-{\psi})D_+{\bar {\psi}})_x+
 D_+(D_+{\psi})^{2},
\nonumber\\
 {\bar \psi}_t &=& {\bar \psi}_{xx}+2D^{-1}_-((D_-{\bar \psi})D_+{\psi})_x+
 D_+(D_+{\bar \psi})^{2};
\label{24}
\end{eqnarray}
\begin{eqnarray}
 {\psi}_t &=& -{\psi}_{xx}+2D_-((D_-{\psi})D_+{\bar {\psi}})+
 D_+(D_+{\psi})^{2},
\nonumber\\
 {\bar \psi}_t &=& {\bar \psi}_{xx}+2D_-((D_-{\bar \psi})D_+{\psi})+
 D_+(D_+{\bar \psi})^{2},
\label{25}
\end{eqnarray}
respectively. Inspection of (\ref{25}) shows that, like in the case
of (\ref{20}), it possesses the $N=2$ supersymmetry, and the $U(1)$
transformation from the $N=2$ supergroup is realized nonlocally.

In terms of ${\psi}$ and ${\bar {\psi}}$, the substitution (\ref{2b}) can
be rewritten in the local form
\begin{equation}
\r {{\bar {\psi}}}{}={\psi}+D_+\ln D_-{\bar {\psi}},\quad
\r {{\psi}}{}={\bar {\psi}}.
\label{26}
\end{equation}
It is instructive to consider its
bosonic limit as well as the bosonic limit of the corresponding
eqs.(\ref{24}). In terms of the bosonic components of the fermionic
superfields ${\psi}$ and ${\bar {\psi}}$, defined as
\begin{equation}
D_+{\psi} \vert={\bar q},
\quad D_-{\psi} \vert={\bar p}, \quad
D_+{\bar {\psi}} \vert=q,\quad D_-{\bar {\psi}} \vert=p ,
\label{boscom1}
\end{equation}
we get the following expressions for the substitution:
\begin{equation}
\r {q}{}={\bar q}+(\ln p)_x,\quad \r {{\bar q}}{}=q, \quad
\r {p}{}={\bar p}-\frac{q_y}{p},\quad \r {{\bar p}}{}=p,
\label{bossub1}
\end{equation}
and for the (1+2)-dimensional bosonic equations:
\begin{eqnarray}
{\bar q}_t &=& -{\bar q}_{xx} + ({\bar q}^{2})_x
             + 2{{\partial}_y}^{-1} ({\bar q}_y q-{\bar p}p_x)_x, \quad
{\bar p}_t = -{\bar p}_{xx}+2({\bar p} q)_x + 2 {\bar p}_x {\bar q},
\nonumber\\
 q_t &=& q_{xx} + (q^{2})_x +
 2{{\partial}_y}^{-1} (q_y{\bar q}-p{\bar p}_x)_x,\quad
 p_t = p_{xx}+2(p{\bar q})_x + 2 p_x q.
\label{bos1}
\end{eqnarray}
One can see from (\ref{bos1}) that at the reduction to the one
dimensional case, the (1+1)-dimensional equations are local. Obviously,
(\ref{bossub1}) and (\ref{bos1}) are integrable because they are obtained
 from integrable supersymmetric counterparts in the bosonic limit.

Equations (\ref{22})-(\ref{25}) and (\ref{bos1}) have complex
structure. It is easy to find the following complex conjugation properties
and scaling dimensions of all superfields and their components:
$b^{*}=-{\bar b}$, ${\bar b}^{*}=-b$, $[b]=[{\bar b}]={cm}^{-1}$;
${\psi}^{*}=-{\bar {\psi}}$, ${\bar {\psi}}^{*}=-{\psi}$,
$[{\psi}]=[{\bar {\psi}}]={cm}^{-1/2}$;
$q^{*}=-{\bar q}$, ${\bar q}^{*}=-q$, $p^{*}=-{\bar p}$, ${\bar p}^{*}=-p$,
$[q]=[{\bar q}]=[p]=[{\bar p}]={cm}^{-1}$. Evolution variable $t$
is purely imaginary and under complex conjugation $t^{*}=-t$.

{}~

{\large {\bf 4.3 Substitution (\ref{6}) }}
\begin{eqnarray}
 v_t &=& -v_{xx}+2(D_+v)D^{-1}_-(uv)_x-
 2vD^{-1}_-\{(vD_+u)_x+2uvD^{-1}_-(uv)_x\},
\nonumber\\
 u_t &=& u_{xx}+2(D_+u)D^{-1}_-(uv)_x-
 2uD^{-1}_-\{(uD_+v)_x - 2uvD^{-1}_-(uv)_x\}.
\label{27}
\end{eqnarray}
At the reduction to the $(1|2)$ superspace, expressions (\ref{27})
are localized and look like
\begin{eqnarray}
 v_t &=& -v_{xx}-
 2iv\{(Dv){\bar D}u-({\bar D}v)Du-\frac{1}{2}v[{\bar D},D]u-i(uv)^{2}\}
-2iu(Dv){\bar D}v,
\nonumber\\
 u_t &=& u_{xx}-
 2iu\{(Du){\bar D}v-({\bar D}u)Dv-\frac{1}{2}u[{\bar D},D]v+i(uv)^{2}\}
-2iv(Du){\bar D}u.
\label{29}
\end{eqnarray}
One can see that the eqs.(\ref{29}) are local and they admit the local
$N=2$ supersymmetry.

As in the previous cases, we now present the bosonic limits of the
substitution (\ref{6}) and the corresponding eqs.(\ref{27}).

Let us define the bosonic components of the bosonic superfields
$u$ and $v$ as
\begin{equation}
v \vert={\bar s},\quad D_-D_+v \vert={\bar r}, \quad
u \vert=s,\quad D_-D_+u \vert=r.
\label{boscom2}
\end{equation}
In terms of these components, (\ref{6}) and (\ref{27}) become
\begin{equation}
 \r {s}{}=\frac{1}{{\bar s}}, \quad
 \r {\bar s}{}={\bar r}-s{\bar s}^{2},\quad
 \r {r}{}=-\frac{{\bar r}}{{\bar s}^{2}},\quad
 \r {\bar r}{}=-{\bar s}(\r {\bar s}{} \r r{} +{\bar s} r
               + s{\bar r} + ({\ln {\bar s}})_{xy});
\label{bossub2}
\end{equation}
\begin{eqnarray}
{\bar r}_t &=& -{\bar r}_{xx} + 2{\bar s}_x(s{\bar s})_x
+2{\bar s}(s_x{\bar s})_x + 2{\bar r}{{\partial}_y}^{-1}(s{\bar r}
-(s{\bar s})^{2})_x-4s{\bar s}^{2}{{\partial}_y}^{-1}({\bar s}r+s{\bar r})_x,
\nonumber\\
 r_t &=& r_{xx} + 2s_x(s{\bar s})_x
+2s(s{\bar s}_x)_x + 2r{{\partial}_y}^{-1}({\bar s}r+(s{\bar s})^{2})_x
+4{\bar s} s^{2}{{\partial}_y}^{-1}({\bar s}r+s{\bar r})_x,
\nonumber\\
{\bar s}_t &=& -{\bar s}_{xx}
-2{\bar s}{{\partial}_y}^{-1}({\bar s}r+(s{\bar s})^{2})_x, \quad
s_t = s_{xx}-2s{{\partial}_y}^{-1}(s{\bar r}-(s{\bar s})^{2})_x,
\label{bos2}
\end{eqnarray}
respectively.

Equations (\ref{27}), (\ref{29}) and (\ref{bos2}) possess the global
$U(1)$-invariance and admit complex structure. With respect to the U(1)
group, the superfields $u$ and $v$ have opposite $U(1)$-charges. Due to the
$U(1)$-invariance, only scaling dimensions of $U(1)$-invariant products are
fixed: $[uv]=[s{\bar s}]={cm}^{-1}$, $[s{\bar r}]=[{\bar s}r]={cm}^{-2}$,
as are the relations among the components with the same $U(1)$ charge,
$[{\bar r}]=[{\bar s}]\times{cm}^{-1}$, $[r]=[s]\times{cm}^{-1}$. The
complex conjugation properties of the superfields and their bosonic
components are the following:  $v^{*}={\pm}iu$, $u^{*}={\pm}iv$;
$s^{*}={\pm}i{\bar s}$, ${\bar s}^{*}={\pm}is$, $r^{*}={\pm}i{\bar r}$,
${\bar r}^{*}={\pm}ir$; $t^{*}=-t$.

{}~

It appears that the $N=2$ supersymmetric integrable systems constructed
here in the $(1|2)$ superspace do not belong to the wide class of $N=2$
supersymmetric integrable hierarchies that have recently been explicitly
constructed using the Lax pair approach in \cite{5}, \cite{6}. It would
be interesting to understand how our hierarchies would be described in
that formalism.

\section{Conclusion}

In this paper, we have applied the formalism of integrable mappings to the
problem of the construction of supersymmetric integrable hierarchies
in the $(2|2)$ and $(1|2)$ superspaces with manifest $N=2$ supersymmetry.
We have proposed three supersymmetric integrable mappings and found
three, to our knowledge, new hierarchies of integrable supersymmetric
systems corresponding to them in the $(2|2)$ superspace possessing
manifest $N=2$ supersymmetry at the reduction to the $(1|2)$ superspace.
New bosonic substitutions and hierarchies are obtained from the
constructed supersymmetric counterparts in the bosonic limit.

We attempted to demonstrate that almost all information about the
integrable hierarchy is encoded in its integrable substitution. In this
context, it seems to be very important to have some reserve integrable
supersymmetric substitutions and integrable hierarchies corresponding to
them in order to better understand their rigorous mathematical structure,
the relations among them and their origins. An analysis of this intriguing
problem is under way.

\section{Acknowledgments}

This work was partially supported by the RFFR grant 96-02-17634, INTAS
grants 93-633, 94-2317, and by a grant from the Dutch NWO organization.

\end{document}